\let\paperbibstyle\bibliographystyle \def\bibliographystyle#1{}
\documentclass[pdflatex,sn-mathphys-num]{sn-jnl}
\let\bibliographystyle\paperbibstyle

\usepackage[T1]{fontenc}
\usepackage{lmodern}
\usepackage{amsmath,amssymb,amsthm,mathtools}
\usepackage{microtype}
\usepackage{xcolor}
\usepackage{graphicx}
\usepackage[bottom]{footmisc}

\hypersetup{colorlinks=true, urlcolor=blue, linkcolor=blue, citecolor=blue,
  pdfauthor={Peixue Wu}}

\usepackage[hyperpageref]{backref}
\renewcommand*{\backref}[1]{}
\renewcommand*{\backrefalt}[4]{%
  \ifcase #1
  \or
    \space(Cited on page~#2)
  \else
    \space(Cited on pages~#2)
  \fi
}
\usepackage{etoolbox}
\makeatletter
\patchcmd{\NAT@citexnum}{\@safe@activesfalse}
  {\@safe@activesfalse\Hy@backout{\@citeb\@extra@b@citeb}}{}
  {\PackageWarning{paper}{natbib/backref patch not applied}}
\makeatother

\newtheorem{theorem}{Theorem}[section]
\newtheorem{lemma}[theorem]{Lemma}
\newtheorem{proposition}[theorem]{Proposition}

\theoremstyle{definition}

\theoremstyle{remark}

\numberwithin{equation}{section}

\DeclareMathOperator{\Tr}{Tr}

\newcommand{\diag}{\mathrm{diag}}
\newcommand{\C}{\mathbb C}

\newcommand{\A}{\mathcal A}
\newcommand{\M}{\mathcal M}

\newcommand{\mb}{\mathbb}
\newcommand{\mc}{\mathcal}
\newcommand{\ket}[1]{\lvert #1\rangle}
\newcommand{\bra}[1]{\langle #1\rvert}

\newcommand{\br}{\boldsymbol{r}}
\newcommand{\by}{\boldsymbol{y}}
\newcommand{\Cchi}{\chi_{\max}}

\definecolor{placeholder}{rgb}{0.48,0.12,0.12}

\begin{document}

\title[Additivity for classical communication]{Classical communication over quantum channels: does entanglement help in nature?}

\author*[1,2]{\fnm{Peixue} \sur{Wu}}

\affil[1]{\orgdiv{Dept. of Mathematics}, \orgname{Syracuse University}, \orgaddress{\city{Syracuse}, \state{NY}, \country{USA}}}

\affil[2]{\orgdiv{Institute for Quantum \& Information Sciences}, \orgname{Syracuse University}, \orgaddress{\city{Syracuse}, \state{NY}, \country{USA}}}


\abstract{Can entangling codewords across independent uses of a quantum
channel improve the rate of classical communication? An old conjecture states that there is no such entanglement advantage for all channels. Although Hastings's result and follow-up works disproved this conjecture, it turns out that enormous dimension is required to construct a channel where entanglement enhances classical communication. This indicates that there might be no entanglement advantage for many physical channels, or for all low dimensional channels. Motivated by recent progress on (generalized) amplitude damping channels, we propose a general Schur-weighted block entropy inequality, and show that there is no entanglement advantage for all phase-covariant channels including multi-level amplitude damping channels, generalized dephasing channels, and damping-dephasing channels. As a by-product, we also recover the recently established result that there is no entanglement advantage for all qubit channels.}


\maketitle
{\renewcommand{\thefootnote}{\fnsymbol{footnote}}%
 \footnotetext[1]{\href{mailto:wpeixue@syr.edu}{wpeixue@syr.edu}}}
\tableofcontents
\section{Introduction}

When independent uses of a quantum channel are employed to transmit classical
information, can entangling the inputs across channel uses increase the
achievable rate? This question is fundamental to quantum Shannon theory, and it
has proved far more subtle than it first appears.

\begin{figure}[htb]
\centering
\includegraphics{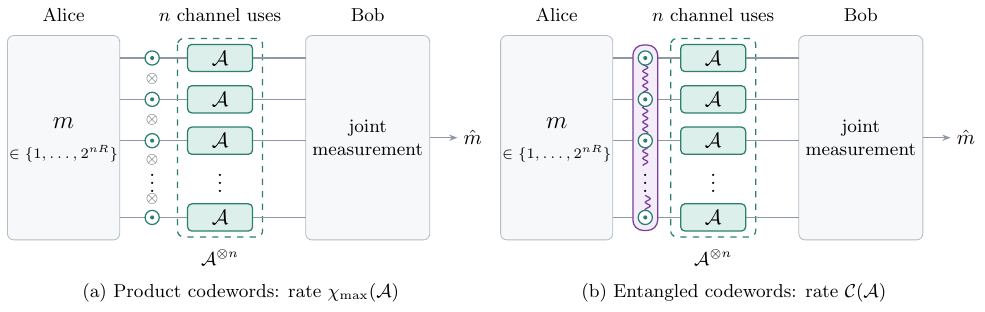}
\caption{Two ways to send a message $m\in\{1,\dots,2^{nR}\}$ over $n$ uses of a
channel $\mathcal A$. (a)~Each codeword is a product state across the channel uses;
the optimal rate is the Holevo information $\Cchi(\mathcal A)$ in~\eqref{eq:holevo}.
(b)~Codewords may be entangled across the channel uses; the optimal rate is the
classical capacity $\mathcal C(\mathcal A)$ in~\eqref{eq:capacity}. In both cases Bob
decodes with a joint measurement of all $n$ outputs.}
\label{fig:codewords}
\end{figure}

For a channel $\mathcal A$, the optimal rate achievable with \textit{product-state} codewords
and collective decoding is the Holevo information
\begin{equation}\label{eq:holevo}
  \Cchi(\mathcal A)
  =\max_{\{p_x,\rho_x\}}\Big[S\big(\mathcal A(\bar\rho)\big)
   -\sum_xp_x\,S\big(\mathcal A(\rho_x)\big)\Big],
\end{equation}
where $\bar\rho=\sum_xp_x\rho_x$, the maximum runs over finite ensembles of
input states, and $S(\rho)=-\Tr\rho\log_2\rho$ is the von Neumann entropy.
Allowing entangled codewords yields the classical
capacity~\cite{Holevo,SWcoding}
\begin{equation}\label{eq:capacity}
  \mathcal C(\mathcal A)
  =\lim_{n\to\infty}\frac1n\,\Cchi(\mathcal A^{\otimes n}).
\end{equation}
The additivity problem asks whether entangled encoding can enhance classical
communication, that is, whether some channel satisfies
$\mathcal C(\mathcal A)>\Cchi(\mathcal A)$. Figure~\ref{fig:codewords} compares the two settings.

For a long time, additivity was expected to hold for all channels. This
expectation was supported by analytical and numerical
studies~\cite{AmosovHolevoWerner2000,KingRuskai2001,OsawaNagaoka2001}, and
additivity was proved for entanglement-breaking~\cite{Shor}, unital
qubit~\cite{KingUnital}, and depolarizing channels~\cite{KingDepol,Amosov},
with extensions to related qudit channels~\cite{DattaRuskai,Fukuda2005} and
approximate versions~\cite{Leditzky_2018}, and more recently generalized amplitude damping channels \cite{Tang2026,Fang2026,OpenAI2026}. Hastings showed that additivity
fails in general~\cite{Hastings}, but all the related constructions require enormous dimensions. This suggests that additivity should
be the typical behavior of low-dimensional channels and many physical channels. 

In this work, motivated by recent progress~\cite{Tang2026,Fang2026,OpenAI2026,Kitamura2026}, we propose a new entropy inequality, named as \textit{Schur-weighted block entropy inequality} (c.f. Theorem~\ref{thm:block-entropy-inequality}), which implies additivity for phase-covariant channels in any dimension (c.f. Theorem~\ref{thm:pc}), and all qubit-to-qubit channels (c.f. Theorem~\ref{thm:qubit-output}). 

\section{Entropic criteria for strong additivity}
We write $\mc A:A\to B$ for a completely positive, trace-preserving map from the input system $A$ to the output system $B$. Unless indicated otherwise, logarithms are to base two. For a unit vector $|\psi\rangle$, the symbol $\psi$ denotes the density matrix $|\psi\rangle\langle\psi|$.

We call $\A$ \emph{strongly Holevo additive}
if for every partner channel $\mc M$,
\begin{equation}\label{eq:strong}
  \Cchi(\A\otimes\M)=\Cchi(\A)+\Cchi(\M).
\end{equation}
In words, entangling the inputs of $\A$ with those of any other channel never increases the Holevo information. Taking $\M=\A^{\otimes(n-1)}$ and inducting on $n$ gives $\Cchi(\A^{\otimes n})=n\,\Cchi(\A)$, and hence
$\mathcal C(\A)=\Cchi(\A)$.

A standard sufficient condition is superadditivity of the convex closure of the output entropy, established by Holevo and Shirokov~\cite[Theorem~1(iv)]{Holevo_2004} specialized to a fixed
pair of channels. We recall this criterion here:
\begin{lemma}\label{lem:entropic-criteria-old}
    Suppose $\mc A: A \to B,\ \mc M: A' \to B'$. If for any bipartite pure state $|\psi\rangle_{AA'}$ with reduced state $\rho_A, \rho_{A'}$, there exist ensembles of states $\{p_\ell,\eta_\ell\}$ on $A$ and $\{q_k,\tau_k\}$ on $A'$, 
    \begin{align}
        & \rho_A = \sum_\ell p_\ell \eta_\ell,\quad \rho_{A'} = \sum_k q_k \tau_k; \label{eq:compatible} \\
        & S(\mc A \otimes \mc M(\psi)) \ge \sum_\ell p_\ell S(\mc A(\eta_\ell)) + \sum_k q_k S(\mc M(\tau_k)), \label{eq:entropic-criteria}
    \end{align}
    then we have $\chi_{\max}(\mc A \otimes \mc M) = \chi_{\max}(\mc A) + \chi_{\max}(\mc M)$.
\end{lemma}
The phase-covariant application will use a weaker compatibility condition
on the first system. Recall that for any optimal ensemble $\{p_\ell,\eta_\ell\}$ for $\chi_{\max}(\mc A)$, the \emph{optimal output center}~\cite[Corollary~1]{Holevo_2004} defined by
\[
\sigma_{\mc A}:= \sum_\ell p_\ell \mc A(\eta_\ell)
\]
is unique, i.e., irrelated to the choice of the optimal ensemble. Every output of $\mc A$ is supported on the support of $\sigma_{\mc A}$. A weaker entropic criteria is then given as follows, which is a direct application of \cite{Holevo_2004}:
\begin{proposition}\label{prop:entropic-criteria-weaker}
     Suppose $\mc A: A \to B,\ \mc M: A' \to B'$. If for any bipartite pure state $|\psi\rangle_{AA'}$ with reduced state $\rho_A, \rho_{A'}$, there exist ensembles of states $\{p_\ell,\eta_\ell\}$ on $A$ and $\{q_k,\tau_k\}$ on $A'$, 
    \begin{align}
        & \sum_\ell p_\ell \Tr(\mc A(\eta_\ell) \log_2 \sigma_{\mc A}) = \Tr(\mc A(\rho_A) \log_2 \sigma_{\mc A}),\quad \rho_{A'} = \sum_k q_k \tau_k; \label{eq:compatible-weaker}\\
        & S(\mc A \otimes \mc M(\psi)) \ge \sum_\ell p_\ell S(\mc A(\eta_\ell)) + \sum_k q_k S(\mc M(\tau_k)),\label{eq:entropic-criteria-weaker}
    \end{align}
    then we have $\chi_{\max}(\mc A \otimes \mc M) = \chi_{\max}(\mc A) + \chi_{\max}(\mc M)$.
\end{proposition}
It is easy to see that \eqref{eq:compatible} implies \eqref{eq:compatible-weaker}, and we will see that it is crucial to weaken the criteria for higher dimensional phase-covariant channels.

These criteria concern a specified pair of channels. Shor's celebrated equivalence theorem~\cite{Shor2004}, by contrast, relates the corresponding additivity and superadditivity conjectures when they are required to hold universally for all channels.

\section{A Schur-weighted block entropy inequality}
The key estimate compares a positive block matrix with a smaller matrix
formed from its block traces. Throughout this section, we extend the entropy to arbitrary positive semidefinite matrices by
\[
S(X)=-\Tr X\log_2 X,
\quad 0\log_2 0:=0.
\]
 
Fix $m\ge1$ and block sizes $n_1,\ldots,n_m\ge1$. Set
$C=(c_{ij})\in\mb M_m$, a block matrix $H=(H_{ij})_{1\le i,j\le m}$ on
$\bigoplus_i\C^{n_i}$, and $W=\bigoplus_iW_i$ with $W_i \ge 0$ acting on $\C^{n_i}$. Define the block Schur multiplication
\begin{equation}
    C\odot H=(c_{ij}H_{ij})_{1\le i,j\le m}.
\end{equation}
We set $\mc P$ as the pinching map on $\bigoplus_i\C^{n_i}$. When $n_i=1$ for all $i$, $\mc P$ keeps the diagonal.
 
For two positive $J,W$ acting on $\bigoplus_i\C^{n_i}$ with $W$ being block diagonal, define the \textit{entropy deficit} $\Delta(\cdot,\cdot)$:
\begin{equation}
    \Delta(J,W) = S(J+W) - S(\mc P(J)) - S(W).
\end{equation}
 
\begin{theorem}[Schur-weighted block entropy inequality]\label{thm:block-entropy-inequality}
Assume $C=(c_{ij})\in\mb M_m$, $H=(H_{ij})_{1\le i,j\le m}$ acting on
$\bigoplus_i\C^{n_i}$, and $W=\bigoplus_iW_i$ with $W_i \ge 0$ acting on $\C^{n_i}$ are all positive semidefinite. Then
\begin{equation}\label{eq:entropy-inequality-block}
    \Delta(C\odot H,W) \ge \Delta(D_{\br}^{1/2} CD_{\br}^{1/2}, D_{\by})
\end{equation}
where $r_i=\Tr H_{ii}$, $y_i=\Tr W_i$, and $D_{\br}=\diag(r_1,\cdots, r_m)$. Equivalently, 
\begin{equation}
    S(C\odot H + W) \ge S(D_{\br}^{1/2}CD_{\br}^{1/2}+D_{\by}) + \sum_{i} c_{ii}r_i S(H_{ii}/r_i) + \sum_i y_i S(W_i/y_i)
\end{equation}
\end{theorem}
Note that this result can cover the technical result \cite[Theorem 4.2]{Tang2026}, see also \cite[Theorem 2.1]{OpenAI2026} by setting $C$ to be all one matrix. To prove the theorem, we first show \eqref{eq:entropy-inequality-block} when $H$ is rank one, i.e., $H = \ket{u}\bra{u}$, where
\begin{equation}
    \ket{u} = \bigoplus_{i=1}^m \ket{u_i},\quad \ket{u_i} \in \mb C^{n_i},\quad \langle u_i|u_i\rangle = r_i.
\end{equation}
Then $H = (\ket{u_i}\bra{u_j})_{1\le i, j \le m}$. Suppose $\{\ket{e_1},\cdots,\ket{e_m}\}$ is the standard basis for $\mb C^m$, define the isometry
\[
V: \mb C^m \to \bigoplus_{i=1}^m \mb C^{n_i},\quad V\ket{e_i} = \frac{\ket{u_i}}{\sqrt{r_i}}
\]
(if $r_i=0$, let $V\ket{e_i}$ be any unit vector of $\C^{n_i}$).
Then it is direct to verify that for $A = D_{\boldsymbol{r}}^{1/2} C D_{\boldsymbol{r}}^{1/2}$ and $J = C\odot H$,
\begin{equation}\label{eq:isometry-relation}
    J = VAV^\dagger,\quad \mc P(J) = V \diag(A) V^\dagger.
\end{equation}
Using invariance under isometry and additivity of von Neumann entropy under direct sum, we have
\begin{equation}\label{eq:entropy-deficit-additivity}
    \begin{aligned}
    S(J + \bigoplus_{i=1}^m y_iI_{n_i}) - S(\bigoplus_{i=1}^m y_iI_{n_i})= S(A + D_{\boldsymbol y})  - S(D_{\boldsymbol{y}}).
\end{aligned}
\end{equation}
We conclude \eqref{eq:entropy-inequality-block} when $H$ is rank $1$ as follows:
\begin{align}
   \Delta(J,W)& = S(J +W) - S(W) - S(\mc P(J)) \\
   & = S(J +W) - S(W) - S(\diag(A)) \label{eq:rank-1-entropy-equality}\\
   & \ge  S(J + \bigoplus_{i=1}^m y_i I_{n_i}) - S(\bigoplus_{i=1}^m y_i I_{n_i}) - S(\diag(A)) \label{eq:rank-1-log-monotone}\\
   & = S(A + D_{\boldsymbol y})  - S(D_{\boldsymbol{y}}) -  S(\diag(A)) = \Delta(A,D_{\boldsymbol{y}}) \label{eq:rank-1-entropy-equality-2},
\end{align}
where \eqref{eq:rank-1-entropy-equality} follows from \eqref{eq:isometry-relation}; \eqref{eq:rank-1-log-monotone} follows from $\log$ is operator monotone thus for $J \ge 0$ and $0\le W \le \bigoplus_{i=1}^m y_i I_{n_i}$, we have
\[
S(J +W) - S(W)  \ge S(J + \bigoplus_{i=1}^m y_i I_{n_i}) - S(\bigoplus_{i=1}^m y_i I_{n_i}).
\]
Finally, \eqref{eq:rank-1-entropy-equality-2} follows from \eqref{eq:entropy-deficit-additivity}. \qed
 
To pass from rank-one matrices to general $H$, we need to combine their
entropy bounds. The entropy deficit is not linear in $H$, so our trick is to first minimize over the block-diagonal $W$ after subtracting a linear functional. The resulting function is concave in the matrix $J \ge 0$, whereas the scalar comparison is jointly convex in its trace parameters. These two concavity/convexity properties allow us to combine the rank-one bounds.

The proof ingredients are matrix maximum principles
and trace-function concavity, see
Lempert~\cite{Lempert2015} and Lieb~\cite{Lieb}, respectively.
Complete Lean proofs of both lemmas are available in the accompanying
repository~\cite{WuLean2026}.
\begin{lemma}\label{lemma:concavity}
    Suppose $\Lambda = \bigoplus_{i=1}^m \mu_i I_{n_i}$ with $\mu_i < 0$. For any $J\ge 0$ acting on $\bigoplus_i\C^{n_i}$, the minimum
    \begin{equation}
        \widehat G_\Lambda(J) = \min_{W\ge 0,\ W\text{ block diagonal}} G_\Lambda(J,W),\quad G_\Lambda(J,W):= \Delta(J,W) - \Tr(\Lambda W)
    \end{equation}
    is attained and $J \mapsto \widehat G_\Lambda(J)$ is concave.
\end{lemma}
 
A jointly convex lemma for scalar function:
\begin{lemma}\label{lemma:convexity}
    For any $C\ge 0$ acting on $\mb C^m$, the function
    \begin{equation}
        f_C(\boldsymbol{r}, \boldsymbol{y}):= \Delta(D_{\br}^{1/2} C D_{\br}^{1/2}, D_{\by})
    \end{equation}
    is jointly convex on $[0,\infty)^{m} \times [0,\infty)^{m}$.
\end{lemma}
 
\begin{proof}[Proof of Theorem~\ref{thm:block-entropy-inequality}]
Denote $r_i=\Tr H_{ii}$, $y_i=\Tr W_i$, and $D_{\br}=\diag(r_1,\cdots, r_m)$. Set \[
A = D_{\br}^{1/2} C D_{\br}^{1/2},\quad J = C\odot H.
\]
Recall $f_C(\boldsymbol{r}, \boldsymbol{y})= \Delta(D_{\br}^{1/2} C D_{\br}^{1/2}, D_{\by})$. Assume first that $C,H,W$ are strictly positive; the general case follows by continuity, since both sides of~\eqref{eq:entropy-inequality-block} are continuous in $(C,H,W)$. Then
\begin{equation}\label{eq:mu-negative}
    \mu_i := \frac{\partial f_C}{\partial y_i}(\br,\by) = \log y_i - [\log(A + D_{\by})]_{ii} < 0.
\end{equation}
Indeed, $\log(A+D_{\by})-\log D_{\by}=\int_0^\infty\big[(D_{\by}+s)^{-1}-(A+D_{\by}+s)^{-1}\big]\,ds$, and the integrand has a positive $i$th diagonal entry because $A_{ii}=c_{ii}r_i>0$. Set $\boldsymbol\mu=(\mu_1,\ldots,\mu_m)$ and $\Lambda=\bigoplus_{i=1}^m\mu_iI_{n_i}$, so that $\Tr(\Lambda W)=\boldsymbol\mu\cdot\by$.
Using convexity of $f_C(\boldsymbol{r}, \boldsymbol{y})$ in the second variable (Lemma~\ref{lemma:convexity}), we have
\begin{equation}\label{eq:convexity-in-y}
    f_C(\boldsymbol{r}, \boldsymbol{z}) \ge f_C(\boldsymbol{r}, \boldsymbol{y}) + \boldsymbol{\mu} \cdot(\boldsymbol{z} - \by),\quad \forall \boldsymbol{z}\in[0,\infty)^m.
\end{equation}
For general $H\ge 0$ acting on $\bigoplus_{i=1}^m \mb C^{n_i}$, the spectral decomposition yields
\begin{equation}
    H = \sum_{\alpha=1}^q p_\alpha \ket{u_\alpha}\bra{u_\alpha},\quad p_\alpha>0,\quad \sum_{\alpha=1}^q p_\alpha=1,
\end{equation}
with $\ket{u_\alpha}=\bigoplus_{i=1}^m \ket{u_{\alpha,i}}$ and $\ket{u_{\alpha,i}}\in\C^{n_i}$. Set \begin{equation}
    r_i^{(\alpha)} = \langle u_{\alpha,i}|u_{\alpha,i}\rangle,\quad J^{(\alpha)} = C \odot \ket{u_\alpha}\bra{u_\alpha},
\end{equation}
so that $\sum_\alpha p_\alpha\br^{(\alpha)}=\br$ and $J=\sum_\alpha p_\alpha J^{(\alpha)}$.
For each $\alpha$, via Lemma~\ref{lemma:concavity}, there exists a minimizer $W^{(\alpha)}$, such that
\[
\widehat G_\Lambda(J^{(\alpha)}) = \Delta(J^{(\alpha)},W^{(\alpha)}) - \Tr(\Lambda W^{(\alpha)}),
\]
and we set $y_i^{(\alpha)}=\Tr W^{(\alpha)}_i$, so that $\Tr(\Lambda W^{(\alpha)})=\boldsymbol\mu\cdot\by^{(\alpha)}$. Then we have
\begin{align}
    \Delta(J,W) - \boldsymbol{\mu} \cdot \by & \ge \widehat G_\Lambda(J) \label{eq:general-definition}\\
    & \ge \sum_\alpha p_\alpha \widehat G_\Lambda(J^{(\alpha)}) \label{eq:general-concavity}\\
    & = \sum_\alpha p_\alpha \left[ \Delta(J^{(\alpha)},W^{(\alpha)}) - \Tr(\Lambda W^{(\alpha)}) \right] \label{eq:general-minimizer}\\
    & \ge \sum_\alpha p_\alpha \left[ \Delta(D_{\br^{(\alpha)}}^{1/2} CD_{\br^{(\alpha)}}^{1/2}, D_{\by^{(\alpha)}}) - \boldsymbol{\mu}\cdot \by^{(\alpha)} \right] \label{eq:general-rank-one}\\
    & = \sum_\alpha p_\alpha \left[ f_C(\br^{(\alpha)}, \by^{(\alpha)}) - \boldsymbol{\mu}\cdot \by^{(\alpha)} \right] \label{eq:general-definition-fC}\\
    & \ge f_C\Big(\br, \sum_\alpha p_\alpha \by^{(\alpha)}\Big) - \boldsymbol{\mu}\cdot  \sum_\alpha p_\alpha \by^{(\alpha)} \label{eq:general-joint-convexity}\\
    & \ge f_C(\br,\by) - \boldsymbol{\mu}\cdot \by, \label{eq:general-convexity-in-y}
\end{align}
where \eqref{eq:general-definition} follows from the definition of $\widehat G_\Lambda(J)$ and $\Tr(\Lambda W)=\boldsymbol\mu\cdot\by$; \eqref{eq:general-concavity} follows from $J=\sum_\alpha p_\alpha J^{(\alpha)}$ and the concavity of $\widehat G_\Lambda$ in Lemma~\ref{lemma:concavity}; \eqref{eq:general-minimizer} follows from the choice of the minimizers $W^{(\alpha)}$; \eqref{eq:general-rank-one} follows from the rank-one case applied to $\ket{u_\alpha}$ and $W^{(\alpha)}$, together with $\Tr(\Lambda W^{(\alpha)})=\boldsymbol\mu\cdot\by^{(\alpha)}$; \eqref{eq:general-definition-fC} follows from the definition of $f_C$; \eqref{eq:general-joint-convexity} follows from the joint convexity of $f_C$ in Lemma~\ref{lemma:convexity} and $\sum_\alpha p_\alpha\br^{(\alpha)}=\br$. Finally, \eqref{eq:general-convexity-in-y} follows from \eqref{eq:convexity-in-y} with $\boldsymbol z=\sum_\alpha p_\alpha\by^{(\alpha)}$.
 
Cancelling $\boldsymbol\mu\cdot\by$ on both ends of the chain gives $$\Delta(J,W)\ge f_C(\br,\by)=\Delta(D_{\br}^{1/2}CD_{\br}^{1/2},D_{\by}),$$ which is~\eqref{eq:entropy-inequality-block}.
\end{proof}

\section{Strong additivity for all phase-covariant channels}\label{sec:pc}

A channel $\mc A: \mb M_d \to \mb M_d$ is called phase-covariant, if it is covariant under every diagonal unitary on a fixed basis: for all $U_{\boldsymbol{\theta}}
=\mathrm{diag}(e^{i\theta_1},\ldots,e^{i\theta_d}),\
\boldsymbol{\theta}\in\mathbb{R}^d.$ \[
\mathcal{A}(U_{\boldsymbol{\theta}}\rho U_{\boldsymbol{\theta}}^{\dagger})
=U_{\boldsymbol{\theta}}\mathcal{A}(\rho)U_{\boldsymbol{\theta}}^{\dagger},\quad \forall \rho.
\]
A normal form of such channels is given in \cite{SinghNechita2021}:
\begin{lemma}[Phase-covariant normal form]\label{lem:pc-normal}
A channel $\mc A: \mb M_d \to \mb M_d$ is phase-covariant on a fixed basis $\{|i\rangle\}_{0\le i \le d-1}$ if and only if 
\begin{equation}\label{eq:pc-normal}
 \A(\rho)
 =C\odot\rho
 +\sum_{i\ne j}p_{ij}\bra j\rho\ket j\,\ket i\bra i,
\end{equation}
where $\odot$ denotes Schur multiplication, i.e., entrywise multiplication in the chosen basis. Here
\begin{equation}\label{eq:pc-cptp}
 p_{ij}\ge0,\quad
 \sum_i p_{ij}=1,\quad
 C=(c_{ij})\succeq0,\quad
 c_{ii}=p_{ii}.
\end{equation}
\end{lemma}

\begin{theorem}\label{thm:pc}
   Suppose $\mc A: \mb M_d \to \mb M_d$ is phase-covariant, then for any channel $\mc M: A' \to B'$, we have $\Cchi(\A \otimes \M) = \Cchi(\A) + \Cchi(\M)$.
\end{theorem}
\begin{proof}
    We verify the weaker entropic criterion Proposition~\ref{prop:entropic-criteria-weaker}. For for bipartite pure state $|\psi\rangle$, denote 
    \[
    |\psi\rangle = \sum_{i=0}^{d-1}\sqrt{r_i} |i\rangle \otimes |v_i\rangle,\quad \sum_i r_i = 1,
    \]
    where the vectors $|v_i\rangle$ have norm one but need not be orthogonal. Assume $\mc A$ takes the form \eqref{eq:pc-normal}, define 
    \begin{equation}
        H_{ij}:= \sqrt{r_ir_j}\mc M(|v_i\rangle \langle v_j|),\quad W_i = \sum_{j\neq i} p_{ij}H_{jj},\quad r_i = \Tr(H_{ii}),\quad y_i = \Tr(W_i).
    \end{equation}
    Then via direct computation, we have 
    \[
    \A \otimes \M(\psi) = C \odot H + W,\quad H = (H_{ij}),\ W = \bigoplus_i W_i.
    \]
    Applying Theorem~\ref{thm:block-entropy-inequality}, and concavity $y_i S(W_i/y_i)\ge\sum_{j\ne i}p_{ij}r_j S(H_{jj}/r_j)$, we have 
    \[
    \begin{aligned}
            &S(\A \otimes \M(\psi)) \\
            & \ge S(D_{\br}^{1/2} CD_{\br}^{1/2}+ D_{\by}) + \sum_{i=0}^{d-1} p_{ii}r_i S(H_{ii}/r_i) + \sum_{i=0}^{d-1} \sum_{j \neq i}p_{ij}r_j S(H_{jj}/r_j) \\
            & = S(D_{\br}^{1/2} CD_{\br}^{1/2}+ D_{\by}) + \sum_{i=0}^{d-1} r_i S(H_{ii}/r_i),
    \end{aligned}
    \]
    where the last equality uses $\sum_{i}p_{ij} = 1$. Choosing $|\psi_{\br}\rangle = \sum_{i=0}^{d-1} \sqrt{r_i} |i\rangle$, and note that for phase-covariant channel $\sigma_{\mc A}$ is diagonal in the fixed basis, we have 
    \begin{align*}
        & \Tr(\mc A(\psi_{\br}) \log \sigma_{\mc A}) = \Tr(\mc A(\rho_A) \log \sigma_{\mc A}),\quad \rho_{A'} = \sum_{i=0}^{d-1} r_i |v_i\rangle \langle v_i|; \\
        & S(\mc A \otimes \mc M(\psi)) \ge S(\mc A(\psi_{\br})) + \sum_i r_i S(\mc M(|v_i\rangle \langle v_i|)),
    \end{align*}
    where the last inequality uses $S(D_{\br}^{1/2} CD_{\br}^{1/2}+ D_{\by}) = S(\mc A (\psi_{\br}))$. Therefore, the weaker entropic criterion in Proposition~\ref{prop:entropic-criteria-weaker} follow and $\mc A$ satisfies strong additivity.
\end{proof}

\section{Strong additivity for all qubit channels}
For a channel $\mc A:\mb M_d\to\mb M_2$, we show that, in a suitable
output basis, it is the sum of an entanglement-breaking operation
with commuting outputs and a completely positive remainder with
at most $d-1$ Kraus operators. Both summands are trace-nonincreasing,
and their sum is trace preserving. The construction follows Choi's
criterion \cite{Choi1975}.

\begin{lemma}\label{lem:qubit-output-kraus}
Let $\mc A:\mb M_d\to\mb M_2$ be a quantum channel. Then there exist an
orthonormal basis $\{\ket0,\ket1\}$ for $\mb C^2$, positive operators
$E_0,E_1\in\mb M_d$, and unnormalized vectors
$\{\ket{a_s},\ket{b_s}\}_{s=1}^{d-1} \subseteq \mb C^d$ such that
\begin{equation}\label{eq:qubit-output-form}
 \mc A(\rho)
 =\sum_{s=1}^{d-1}K_s\rho K_s^\dagger
  +\sum_{i=0}^{1}\Tr(E_i\rho)\ket i\bra i,
 \quad
 K_s=\ket0\bra{a_s}+\ket1\bra{b_s}.
\end{equation}
The vectors may be zero, and
\begin{equation}\label{eq:qubit-output-coherent-kraus}
 \sum_{s=1}^{d-1}
 \bigl(\ket{a_s}\bra{a_s}+\ket{b_s}\bra{b_s}\bigr)
 +E_0+E_1=I_d.
\end{equation}
\end{lemma}

\begin{proof}
Denote $\mc A^*: \mb M_2 \to \mb M_d$ as the channel in the Heisenberg picture, i.e., the unital and completely positive map dual to $\mc A$. Write $Q_{ij}=\mc A^*(\ket i\bra j)$ where $|0\rangle,|1\rangle$ are basis vectors for $\mb C^2$ to be determinined. Choi's criterion and unitality give
\begin{equation}\label{eq:qubit-output-dual-choi}
 Q=
 \begin{pmatrix}
 Q_{00}&Q_{01}\\
 Q_{01}^\dagger&Q_{11}
 \end{pmatrix}
 \ge0,
 \quad
 Q_{00}+Q_{11}=I_d.
\end{equation}
We choose the output basis $|0\rangle, |1\rangle$ so that
$r:=\operatorname{rank}Q_{01}\le d-1$. In fact, assume
\[
 |0\rangle=\binom{\alpha}{\beta},
 \quad
 |1\rangle=\binom{-\overline\beta}{\overline\alpha}.
\]
Then $\det Q_{01}$ is a polynomial with variables $\alpha,\beta$ and their complex conjugates. $\det Q_{01} = 0$ always has solutions, and for this choice of basis, we have $r \le d-1$. We use them as the output basis from now on.

We now construct the vectors $|a_s\rangle,|b_s\rangle$ and the two positive operators $E_0, E_1$.
If $r=0$, take $a_s=b_s=0$ for all $s$ and $E_i=Q_{ii}$.
Assume henceforth that $r>0$. Denote the spectral decomposition of $Q_{00}$ as 
\[Q_{00}
 =\sum_{j=1}^{k}\mu_j\ket{e_j}\bra{e_j},
 \quad \mu_j>0,\]
Positivity of $Q$ implies $\operatorname{ran} Q_{01}
 \subseteq\operatorname{ran}Q_{00}$, thus we can define
\[
 T=\left( \sum_{j=1}^{k}\mu_j^{-1/2}\ket{e_j}\bra{e_j} \right)Q_{01}.
\]
Take the singular value decomposition (note that $\operatorname{rank} T = \operatorname{rank}Q_{01} = r$)
\begin{equation}\label{eq:qubit-output-rank-factorization}
 T=\sum_{s=1}^{r}\lambda_s\ket{u_s}\bra{v_s},
 \quad \lambda_s>0,
\end{equation}
where $\{u_s\}_{s=1}^{r}$ and $\{v_s\}_{s=1}^{r}$ are
orthonormal vectors with norm $1$. Define
\begin{equation}\label{eq:qubit-output-vector-construction}
 \ket{a_s}=Q_{00}^{1/2}\ket{u_s},
 \quad
 \ket{b_s}=\lambda_s\ket{v_s},
 \quad 1\le s\le r.
\end{equation}
Set $a_s=b_s=0$ for $r<s\le d-1$, and put
\begin{equation}\label{eq:qubit-output-diagonal-remainders}
 E_0=Q_{00}-\sum_{s=1}^{r}\ket{a_s}\bra{a_s},
 \quad
 E_1=Q_{11}-\sum_{s=1}^{r}\ket{b_s}\bra{b_s}.
\end{equation}
These choices reproduce the channel $\mc A$ in the Kraus form~\eqref{eq:qubit-output-form}.
\end{proof}
We now prove strong additivity for channels with one coherent Kraus operator.
\begin{theorem}\label{thm:qubit-output}
Suppose there exist orthonormal basis $\{\ket0,\ket1\}$ for $\mb C^2$, positive operators $E_0,E_1\in\mb M_d$, and unnormalized vectors
$|a_0\rangle, |a_1\rangle \in \mb C^d$ such that
\begin{equation}\label{eq:qubit-output-form-one}
 \mc A(\rho)
 = K \rho K^\dagger
  +\sum_{i=0}^{1}\Tr(E_i\rho)\ket i\bra i,
 \quad
 K=\ket0\bra{a_0}+\ket1\bra{a_1}, 
\end{equation}
where $\ket{a_0}\bra{a_0}+\ket{a_1}\bra{a_1} +E_0+E_1=I_d$. Then for any channel $\mc M: A'\to B'$, we have $\chi_{\max}(\mc A \otimes \mc M) = \chi_{\max}(\mc A) + \chi_{\max}(\mc M)$. In particular, any quantum channel $\mc A: \mb M_2 \to \mb M_2$ satisfies strong additivity.
\end{theorem}
\begin{proof}
    We verify the entropic criteria \eqref{eq:compatible} and \eqref{eq:entropic-criteria} in Lemma~\ref{lem:entropic-criteria-old}. For any bipartite pure state $|\psi\rangle$, denote
    \begin{equation}\label{eq:ensemble-qubit-A'}
        |v_i\rangle := (\langle a_i| \otimes I_{A'})|\psi\rangle \in A',\quad \tau_i := \Tr_A((E_i \otimes I_{A'}) \psi)
    \end{equation}
    with $\sum_{i=0}^1 (|v_i\rangle \langle v_i| + \tau_i)$ being trace one since $\sum_{i=0}^1 (|a_i\rangle \langle a_i| + E_i)=I_d$, and
    \begin{equation}
        H_{ij} := \mc M(|v_i\rangle \langle v_j|),\quad W_i := \mc M(\tau_i).
    \end{equation}
    By definition of $\mc A$ in~\eqref{eq:qubit-output-form-one},
    \begin{align*}
        (\mc A\otimes \mc M)(\psi) &= \sum_{i,j=0}^1 |i\rangle \langle j| \otimes H_{ij} + \sum_{i=0}^1 |i\rangle \langle i| \otimes W_i \\
        & = C \odot \begin{pmatrix}
            H_{00} & H_{01} \\
            H_{01}^\dagger & H_{11}
        \end{pmatrix} + \begin{pmatrix}
            W_0 & 0 \\
            0 & W_1
        \end{pmatrix}
    \end{align*}
    where $C = \begin{pmatrix}
            1 & 1 \\
            1 & 1
        \end{pmatrix}$. Applying Theorem~\ref{thm:block-entropy-inequality}, we have
    \begin{equation}\label{eq:key-step-qubit}
        \begin{aligned}
               & S\left((\mc A\otimes \mc M)(\psi)\right) \\
               & \ge S(D_{\br}^{1/2} C D_{\br}^{1/2} + D_{\by}) + \sum_{i=0}^1 r_i S(H_{ii}/r_i) + \sum_{i=0}^1 y_i S(W_i/y_i),
        \end{aligned}
    \end{equation}
    where \[
    r_i = \Tr(H_{ii}) = \langle a_i| \rho_{A} |a_i\rangle,\quad y_i = \Tr(W_i) = \Tr(E_i \rho_A).
    \]
    Now we show that there exists a pure state ensemble $\{p_\ell, \eta_\ell\}$ with $\rho_{A} = \sum_\ell p_\ell \eta_\ell$, such that $\sum_\ell p_\ell U_\ell \mc A(\eta_\ell) U_\ell^\dagger = D_{\br}^{1/2} C D_{\br}^{1/2} + D_{\by}$ for some unitary operators $U_\ell$, thus by concavity of von Neumann entropy
    \[
    S(D_{\br}^{1/2} C D_{\br}^{1/2} + D_{\by}) \ge \sum_\ell p_\ell S(\mc A(\eta_\ell)).
    \]
    In fact, denote $m = \operatorname{rank}(\rho_A)$ and choose the eigenbasis $\{|e_k\rangle\}$ for traceless Hermitian operator \[
    X:= \rho_A^{1/2}\big[r_1 |a_0\rangle \langle a_0| - r_0 |a_1\rangle \langle a_1| \big] \rho_A^{1/2},
    \]
    and $|u_\ell\rangle = \frac{1}{\sqrt{m}} \sum_{k=0}^{m-1} \exp(\frac{2\pi \mathrm{i} k\ell}{m})|e_k\rangle$. Then we can show that the ensemble
    \begin{equation}
        p_\ell = \langle u_\ell| \rho_A |u_\ell\rangle,\quad |\eta_\ell\rangle = \frac{\rho_A^{1/2} |u_\ell\rangle}{\sqrt{p_\ell}}
    \end{equation}
    meets the requirement. In summary, recalling \eqref{eq:key-step-qubit}, we have \begin{equation}
        \rho_A = \sum_\ell p_\ell \eta_\ell,\quad \rho_{A'} = \sum_{i=0}^1 \left(r_i \frac{|v_i\rangle \langle v_i|}{r_i} + y_i\frac{\tau_i}{y_i}\right)
    \end{equation}
    which verifies \eqref{eq:compatible} in Lemma~\ref{lem:entropic-criteria-old}; \eqref{eq:entropic-criteria} follows from \eqref{eq:key-step-qubit} and $S(D_{\br}^{1/2} C D_{\br}^{1/2} + D_{\by}) \ge \sum_\ell p_\ell S(\mc A(\eta_\ell))$.

    When $d=2$, via Lemma~\ref{lem:qubit-output-kraus}, $\mc A$ can be represented as \eqref{eq:qubit-output-form-one} thus qubit-to-qubit channels satisfy strong additivity.
\end{proof}
\section{Discussions}
In Theorem~\ref{thm:qubit-output}, we did not provide the proof of strong additivity for channels $\mc A:\mb M_d \to \mb M_2$ for any $d >2$, the technical obstruction is when there are more than one coherent Kraus operators (the $K$ term has at most $d-1$, see Lemma~\ref{lem:qubit-output-kraus}), our general entropy inequality does not apply directly. 

Phase-covariance in this work is for all diagonal unitaries, does other type of covariance also imply strong additivity? For example, we can study covariant with respect to unitaries given by Hamiltonians $H$: $U_t = \exp(itH), \forall t \in \mb R$.

Our analysis can be a starting point to resolve the following \textit{open problem} on no entanglement advantage for all low dimensional channels: find $N \ge 2$, such that for all channels $\mc A: \mb M_d \to \mb M_d$ with $d \le N$, we have $\mc C(\mc A) = \Cchi(\mc A)$ and if $d > N$, there exists an explicit channel $\mc A: \mb M_d \to \mb M_d$ such that $\mc C(\mc A) > \Cchi(\mc A).$

On a practical level, if we limit the number of uses of channels, can we achieve entanglement advantage? See related discussions in \cite{WildeErasure2026}.

\backmatter

\section*{AI usage}
During the preparation of this manuscript, the author used ChatGPT-6 Astra
and Claude Opus 5.5 to study the OpenAI manuscript~\cite{OpenAI2026} and to explore generalizations and proofs. All the results in this paper are verified by the author and also fully verified in Lean ~\cite{WuLean2026}. The author takes full responsibility for the results and their interpretation.

\bibliographystyle{marcotomPB-appearance}
\bibliography{ref}
\end{document}